\documentclass[pre,twocolumn]{revtex4}
\usepackage{graphicx}
\begin{document}


\newcommand{\laplacian}{\nabla^2}

\newcommand{\changed}[1]{\textbf{#1}}
\newcommand{\remark}[1]{\textsc{#1}}

\title{Saffman-Taylor streamers: mutual finger interaction in an electric breakdown}
\author{Alejandro Luque$^1$, Fabian Brau$^1$ and Ute Ebert$^{1,2}$}
\affiliation{$^1$~CWI, P.O. Box 94079, 1090 GB Amsterdam, The Netherlands}
\affiliation{$^2$~Department of Physics, Eindhoven University of Technology, The Netherlands}

\date{\today}

\begin{abstract}
%
Bunches of streamers form the early stages of sparks and lightning but theory presently concentrates on
single streamers or on coarse approximations of whole breakdown trees. Here a periodic array of interacting
streamer discharges in a strong homogeneous electric field is studied in PDE approximation in two dimensions. If the period of the streamer array is small enough, the streamers do not branch, but approach uniform translation. When the streamers are close to the branching regime, the enhanced field at the tip of the streamer is close to $2 E_{\infty}$, where $E_{\infty}$ is the homogeneous field applied between the electrodes. We discuss a moving boundary approximation to the set of PDEs. This moving boundary model turns out to be essentially the same as the one for two-fluid Hele-Shaw flows. In two dimensions, this model possesses a known analytical solution. The shape of the 2D interacting streamers in uniform motion obtained from the PDE simulations is actually well fitted by the analytically known ``selected Saffman-Taylor finger''. This finding helps to understand streamer interactions and raises new questions on the general theory of finger selection in moving boundary problems.
\end{abstract}

\maketitle

\section{Introduction}
\label{sec:intro}

Streamers are growing ionized fingers that appear in electric breakdown whenever non-ionized matter is
suddenly exposed to strong electric fields, therefore they are very common in nature and technology in
gases, liquids and solids~\cite{raiz91,veld99,eber06}. They occur for instance in early stages of
atmospheric discharges such as sparks and lightning or in sprite discharges high above
thunderclouds~\cite{sent95,pask98,Gerken,pask02,niel07}. Streamers are characterized by a thin space charge
layer around their tip that enhances the local electric field; this enhanced field in turn creates a very
active impact ionization region.

\begin{figure}
\includegraphics[width=\columnwidth]{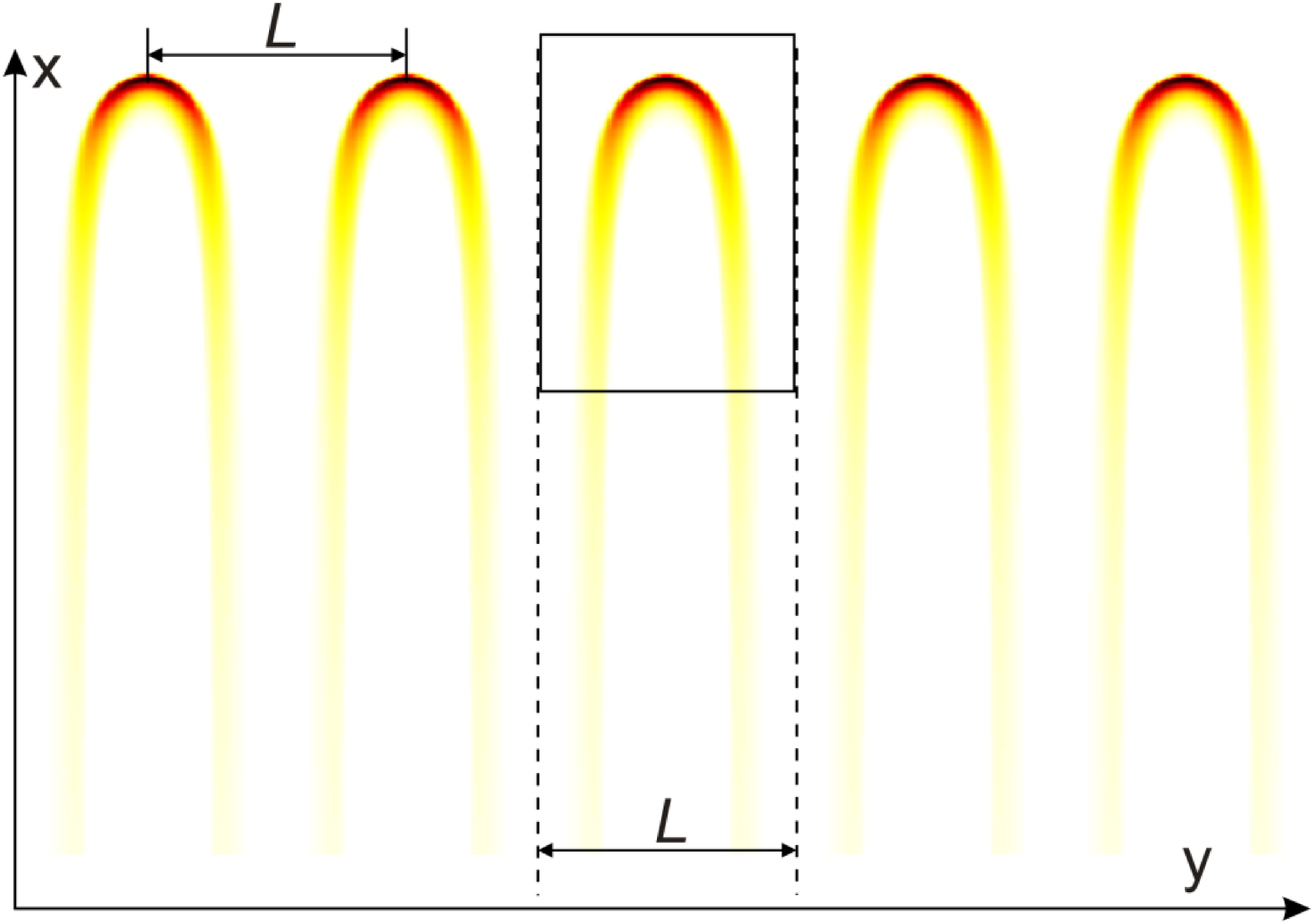}
\caption{Periodic array of negative streamers (net charge density) in a strong homogeneous background electric field $E_{\infty}$ pointing downwards. $L$ is the period of the array. The dash lines represent two symmetry lines. The box around a part of the central streamer indicates the part presented in Fig.~\ref{chargedens}.\label{arraystream}}
\end{figure}

Most experiments produce many streamers, certainly when the emitting electrode is a long wire \cite{Winands06},
and frequently also when it is the point of a needle \cite{Briels06}. Simulations, on the other hand, concentrate
almost exclusively either on single streamers within a microscopic discharge model,
or on the complete streamer branching tree in quite coarse phenomenological
models. Only in~\cite{naid96}, the electrostatic interaction of narrow streamers
within a widely spaced streamer array is studied in relatively low electric fields
within a microscopic model;
as the streamer radius is fixed, the numerical implementation is essentially one-dimensional.
In the present paper, we mimic a similar periodic array
of identical parallel streamers, but in a higher field, see Fig.~\ref{arraystream}. Furthermore, rather than fixing radius and shape of the streamers a priori, we let it emerge dynamically within the simulation.
Such arrays of streamers can be created experimentally by an array of needles
inserted into a plate electrode~\cite{takak05,stari05}. Bunches of parallel streamers have also been
observed in sprite discharges above thunderclouds~\cite{Gerken,niel07}.

The problems addressed in this paper are multiple: What is the charge distribution and velocity of an array
of streamers, depending on their distance and on the applied electric field? Do they approach a state of
uniform translation, in contrast to single streamers? And how can the dynamical evolution of their shape be
placed in the context of other moving boundary problems in nature? Giving already a major conclusion of the
paper, we find that uniformly translating streamer arrays in the microscopic discharge model  in two
spatial dimensions are very well fitted by a classical solution of two-fluid-flow~\cite{pelc88}, namely by
the so-called selected Saffman-Taylor finger~\cite{saff58}, cf.\ Fig.~\ref{chargedens}. Therefore the
velocity of the streamer array is about twice the electron drift motion in the background field, and their
diameter approaches half the period of the array. The observation also raises a theoretical question on pattern
selection, namely why the same finger shape is selected in the hydrodynamic and in the discharge problem,
given the fact that the problems are similar but not identical.

The paper is organized as follows: Sections \ref{sec:model-a} and \ref{sec:model-b} introduce the minimal PDE model for streamers adapted to describe the evolution of an array of streamers. The general behaviour and properties of these interacting streamers are discussed in Sec.~\ref{sec:model-c}. Section \ref{sec:profile-a} presents a moving boundary approximation of the minimal model used in the simulations.
In Sec.~\ref{sec:profile-b} we present
an analytical solution of this approximation for the shape of the streamer. This solution is known as the selected Saffman-Taylor finger and fits well the charge distribution of the streamer, cf. Fig.~\ref{chargedens}.  In Sec.~\ref{sec:profile-c} we briefly discuss some open issues related to this boundary observation.
We conclude by shortly summarizing and discussing our study in Sec.~\ref{sec:conclusions}.

\section{Density approximation and simulation results}
\label{sec:model}

\subsection{Minimal streamer model}
\label{sec:model-a}

We analyze negative streamers in simple media like pure nitrogen within the minimal streamer
model~\cite{raiz91,Kunhardt,dhal85,vite94,eber96,baze98} that includes electron diffusion and drift in a
self-consistent electric field, while ions are taken as immobile due to their much larger mass.  New charge
carriers are generated by an impact ionization term in Townsend approximation that depends nonlinearly on the
local electric field. In dimensionless units, the model is
\begin{eqnarray}
\label{pde1}
\partial_t \sigma &=& D \laplacian \sigma + \mathbf{\nabla} \cdot (\sigma {\bf E})
+ \sigma\, |{\bf E}|\, e^{-1/|{\bf E}|}, \\
\label{pde2}
\partial_t \rho &=& \sigma\, |{\bf E}|\, e^{-1/|{\bf E}|}, \\
\label{pde3}
\laplacian \phi &=& \sigma-\rho, \quad {\bf E}= -\mathbf{\nabla} \phi,
\end{eqnarray}
where $\sigma$ and $\rho$ are the electron and ion densities, $\phi$ is the electrostatic potential, ${\bf
E}$ the electric field, and $D$ is a diffusion coefficient, taken as $D=0.1$~\cite{eber96,eber06}. The
intrinsic length scale of the model is the mean free path of an electron between two ionizing collisions in
fields $|{\bf E}|\gg1$, for nitrogen at standard temperature and pressure, it is $2.3\,\mu\mathrm{m}$;
the scale of time is 3 ps and the scale of the electric field is $\approx 200$~kV/cm in this case. A general
discussion of dimensions can be found, e.g., in~\cite{eber96,eber06,luque07}.  There it is argued that the main advantage of working with dimensionless quantities is that all basic results are immediately generalized to any gas pressure, temperature and composition.

The model is solved numerically
on adaptively refined comoving grids as described in detail in~\cite{mon2006}; the finest grid in our
simulations was 1/4. 

Notice that the only ionization source in our model is impact ionization.  We 
assumed this for the sake of simplicity and in order to emphasize the
elementary processes that are common between streamers and two-phase 
hydrodynamic systems, as will be discussed below.  In exchange for this
simplicity we restricted ourselves to negative streamers in media where 
photo-ionization is absent or negligible, such as pure nitrogen, 
argon or GaAs.  Moreover, in \cite{luque07} it was shown that under certain conditions the effects of photo-ionization on the propagation of negative streamers are negligible even in ambient air.

\subsection{Implementing an array of streamers}
\label{sec:model-b}

Another simplifying assumption is the restriction of the problem to 
two-dimensions. Indeed, we believe that the main characteristics of the dynamics of the interacting streamers in two-dimensions, as described below, will be qualitatively the same as those in three-dimensions. This is supported by past simulations of single streamers in two-dimensions \cite{brau07} which are qualitatively very similar
to three-dimensional simulations \cite{luque07}.
Moreover the generalization to a three-dimensional geometry is not straightforward since the numerical implementation of streamers in a 2D periodic lattice is non-trivial due mainly to the implementation of the boundary conditions. Another reason to consider only a two-dimensional geometry is that it allows the construction of moving boundary approximations with a
known analytical solution. The agreement between that
explicit solution and the actual shape of the front is remarkable and
detailed below. However, two-dimensional streamers are not just
interesting from an academic point of view, they also occur in
experiments in thin semiconductor wafers~\cite{Schoell}.

An infinite, periodic array of streamers can be reduced to the
simulation of a single streamer in a channel with Neumann conditions on the lateral boundaries. This is done
as follows. If the streamer is centered at $y=0$ and propagates along the $x$ direction, and if the period of
the streamer array is $L$, there are two symmetry lines at $y=\pm L/2$ where all normal
derivatives vanish; therefore Neumann conditions for potential and electron density $\partial_y \phi=0$,
$\partial_y \sigma=0$, at $y=\pm L/2$ can substitute the other streamers in the array.

This array of streamers is now studied in a constant electric field ${\bf E}=-E_{\infty}\hat{\bf x}$ far ahead
of the streamers; this field is imposed as an inhomogeneous Neumann boundary condition on $\phi$ at the
boundary at $x\gg1 $ while at $x = 0$ the electrostatic potential is fixed. These conditions can be used
when planar electrodes are first charged and then insulated; for more general electric circuits, they also
approximate streamers that are much shorter than the inter-electrode distance. For the particle densities,
we used homogeneous Neumann boundary conditions on all boundaries. As initial conditions we used in this
paper an electrically neutral Gaussian seed centered at $(x, y) = (0, 0)$ of width $16$ and height $1/4.7$,
except in Fig.~\ref{fronts}b, where the width is larger but the total number of particles is not changed. We verified that the same attractor of the dynamics is approached after sufficiently long time when lateral position, width and height of the initial seed were varied.

\subsection{Branching versus uniform translation}
\label{sec:model-c}

\begin{figure}
\includegraphics[width=\columnwidth]{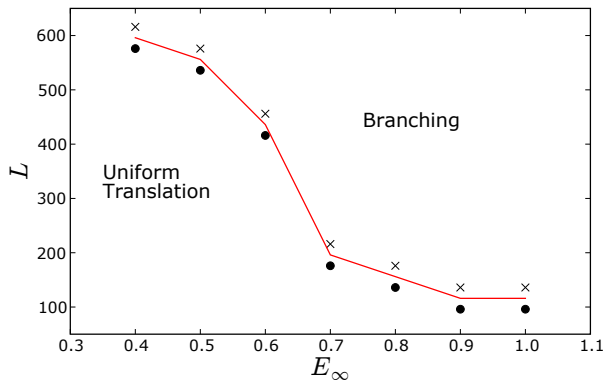}
\caption{Simulated streamers that branch ($\times$) or translate uniformly ($\bullet$) as a function of the
period $L$ of the array and of the uniform field $E_{\infty}$ ahead of it. The line interpolates the phase
transition. \label{phasediagram}}
\end{figure}

After a transient evolution, the simulated streamers either reach a state of uniform translation, i.e. they propagate with constant velocity and unchanged shape, or they
branch like single streamers~\cite{arra02,monbr}. Two parameters control the two regimes for the evolution of the streamers: the period of the array, $L$, and the background electric field applied between the electrodes, $E_{\infty}$. Fig.~\ref{phasediagram} shows a phase diagram spanned by the electric field $E_{\infty}$ and the spatial period $L$; here $L=96$ to $616$ was explored in steps of $\Delta L=40$ and $E_{\infty}=0.4$ to $1.0$ in steps of $\Delta E_{\infty}=0.1$. Below the transition line,
i.e., for small period $L$, the proximity of the other streamers suppresses branching and the whole streamer
array propagates uniformly after some transient stage, while above the line the streamers branch eventually. We remark that in general, there can be uniformly translating solutions in the part of the phase diagram marked as ``branching'';
however, the set of initial conditions for which those solutions emerge (their basin of attraction) is so small that they are not reached from our initial conditions.

\begin{figure}
\includegraphics[width=\columnwidth]{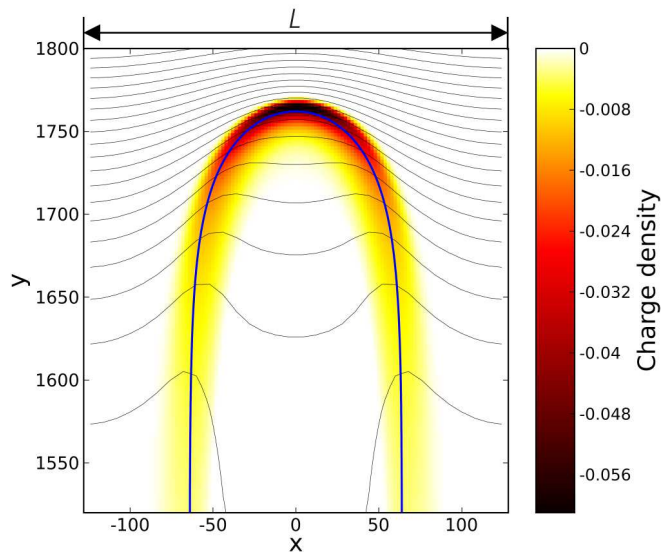}
\caption{(Color online) The thin space charge layer $\rho-\sigma$ around the uniformly translating streamer
discharge (density color coded) with the Saffman-Taylor finger of width $L/2$ superimposed (thick solid line). $L$
is the width of the Hele-Shaw cell for the Saffman-Taylor finger or the period of the array for streamers,
the lateral boundaries then being lines of mirror symmetry between the streamers. Here $L=256$, and the
electric field far ahead is $E_{\infty}=0.5$; this corresponds to $\simeq$~0.059 cm and $\simeq$~100 kV/cm for
nitrogen under normal conditions. Equipotential lines are also plotted (thin solid lines). \label{chargedens}}
\end{figure}

\begin{figure}
\includegraphics[width=\columnwidth]{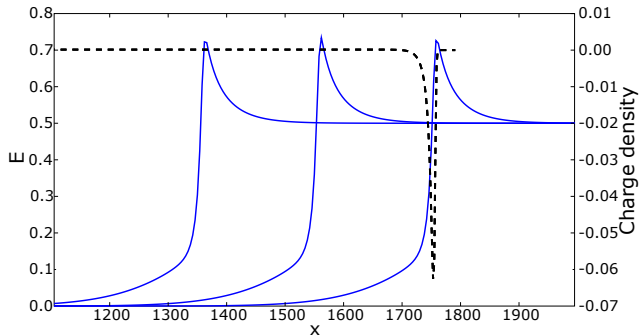}
\caption{Absolute value of the electric field (solid lines) for times $t=1400$, $1600$, and $1800$,
and space charge density (dotted line) at time $t=1800$ on the streamer axis for $L=256$ and
$E_{\infty}=0.5$. Field and density at time $t=1800$ correspond to the uniformly translating finger
in Fig.~\ref{chargedens}. \label{profile}}
\end{figure}

We now analyze in detail the uniformly translating streamer array that emerges for sufficiently small
$E_{\infty}$ and/or $L$ (the lower part of the phase diagram, see Fig.~\ref{phasediagram}). After initial transients of duration $t\cong 100$ or less, these streamer heads reach
a constant velocity and a constant shape for the rest of the evolution: this is the attractor of the dynamics, namely, the solution reached after a sufficiently long evolution from a large set of initial conditions. Therefore this attractor does not depend on the particular choice of the initial seed used. Only the transient evolution and its duration can depend on this choice. But all these various transient regimes lead to the same final uniformly translating state, with the same shape and velocity. Fig.~\ref{chargedens} shows the space charge distribution $\rho-\sigma$ of the attractor for $E_{\infty}=0.5$ and $L=256$ at the time $t=1800$ (long after the transient evolution ended). Figure~\ref{profile} shows the electric field and the net charge profiles from the same simulation. The electric field along the streamer axis ($y=0$) is presented at times 1400, 1600 and 1800 together with the net charge density at time 1800. The three profiles of the electric field show that the
propagation indeed is uniform. The thin space charge layer creates a strong field enhancement immediately
ahead of the ionization fronts like in a freely propagating streamer. However, behind the space charge layer,
the electric field profile inside the streamer array shows characteristic differences to the field profile
within a single streamer~\cite{arra02,monbr,mon2006,luque07,brau07}. Immediately behind the space charge
layer, the electric field decays very rapidly like in a single streamer. Then a transition to a slower field
decay sets in. Finally, far behind the streamer head, the electric field vanishes completely, in contrast to
the nonvanishing residual field inside a single streamer. These observations require further studies.
However, one conclusion can already be drawn by applying the Poisson equation $\nabla\cdot{\bf
E}=\rho-\sigma$ to the streamer head front as a whole. As the field has a constant value $-E_{\infty}$ far
ahead of the streamer array and vanishes far behind the streamer heads, the streamer heads must carry an
average charge $-E_{\infty}$ per unit area, i.e., each streamer head must carry a total charge overshoot of
$-E_{\infty}\cdot L$ to collectively screen the electric field completely behind the array of heads.

These properties of an array of streamers contrast strongly with those of a
single streamer, discussed extensively in \cite{brau07}.
For example, a single streamer in a strong
homogeneous background electric field never reaches a state of uniform translation. The
radius of curvature of the head of the streamer expands during its
motion up to the time where instabilities grow and branching
occurs. Furthermore, the electric field inside of a single streamer
is not as perfectly screened. There are, therefore, remarkable qualitative changes in the
propagation of a streamer when the interaction of neighbouring
streamers is significant. These effects are expected to persist also
in the three-dimensional case.

However, global considerations on the charge content of the streamer head do not fix the shape of the finger and the spatial charge distribution
within each uniformly translating streamer head. These density distributions and the consecutive field
enhancement and velocity are problems of dynamical selection that will be addressed in the remainder of the paper.

\section{Moving boundary approximation and Saffman-Taylor solution}
\label{sec:profile}

\subsection{Moving boundary approximation}
\label{sec:profile-a}

As shown in Figs.~\ref{arraystream} and \ref{chargedens}, after a
sufficiently long evolution, during the steady evolution of the streamer, the width of the ionization front can be much smaller than its radius of curvature. Similarly to other pattern forming systems, such as solidification
fronts, this separation of scales enables one to consider the front as an infinitesimally thin, sharp moving interface. The original nonlinear dynamics is then replaced by a set of linear field equations (typically Laplace)
on both sides of the interface, with appropriate boundary conditions at the interface and further away from it. The interface dynamics is then typically related to gradients of the Laplacian fields at its vicinity.

\begin{figure}
\includegraphics[width=\columnwidth]{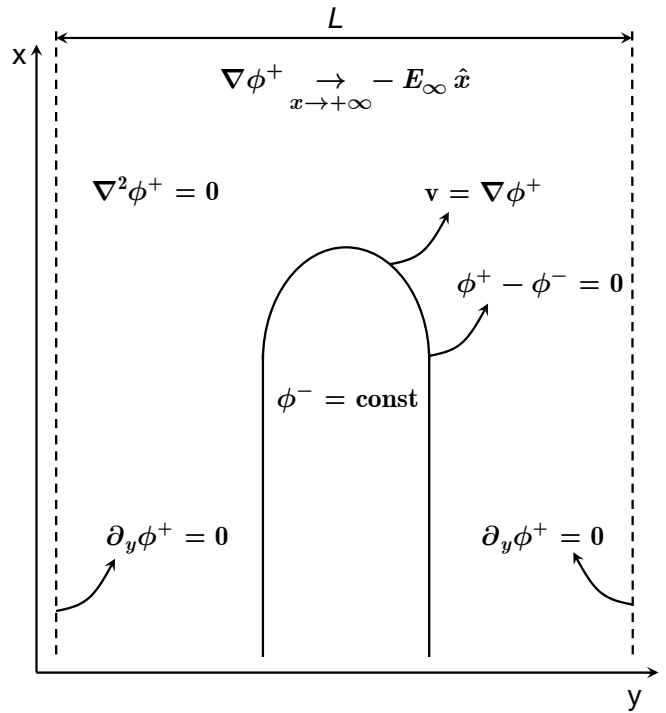}
\caption{Schematic view of the mathematical setup for the moving boundary approximation. $\phi^+$ and $\phi^-$ stand for the electric potential outside and inside the streamer respectively.
\label{mathsetup}}
\end{figure}

In~\cite{arra02,meul04,meul05,eber07,brau07}, a moving boundary approximation was proposed and elaborated for
the thin space charge layer and associated sharp ionization front that separates the ionized from the
non-ionized region. This model was proposed to describe the evolution of a single streamer but it is straightforward to adapt it to the evolution of an array of streamers since we just need to add homogeneous Neumann boundary conditions on the symmetry lines. See Fig.~\ref{mathsetup} for a schematic view of the mathematical setup of the moving boundary approximation. 

The non-ionized, electrically neutral region outside the streamer is fully described by $\nabla^2\phi=0$, and $\phi\to\phi_0+E_{\infty} x$ fixes the homogeneous field $E_{\infty}$ far ahead of the streamers at $x\gg1$. The
symmetry line between two streamers is represented by a Neumann boundary condition for the electric
potential, $\partial_y \phi=0$, at $y=\pm L/2$. If the boundary motion is approximated by the local electron
drift velocity ${\bf v}=\nabla\phi$, the interior of the streamer as ideally conducting $\phi=\text{const.}$
(where the constant can be set to 0 due to electrostatic gauge invariance), and the electric potential across
the boundary as continuous, we arrive precisely at the unregularized moving boundary problem for a
Saffman-Taylor finger after simply substituting the electric potential $\phi$ by the pressure field $p$. This
is a classical problem where a very viscous fluid is penetrated by a much less viscous one within the narrow
spacing of a Hele-Shaw cell. 

\subsection{Comparison with the Saffman-Taylor solution}
\label{sec:profile-b}

An explicit uniformly translating solution for this moving boundary problem was found long ago by Saffman and
Taylor \cite{saff58}. The solution for the interface $x=x(y,t)$ in a channel of width $L$ is given by
\begin{equation}
    \label{steq}
    x=\frac{L(1-\lambda)}{2\pi}\ln\left[\frac12\left(1+\cos\left(\frac{2\pi y}{\lambda L}\right)\right)\right]+v t,
\end{equation}
where the velocity is $v=E_{\infty}/\lambda$ in our notation and the field at the tip is enhanced by a factor $1/\lambda$. The parameter $\lambda$ is the ratio between the
width of the finger and the width of the channel; $\lambda$ can take any value between $0$ and $1$,
parametrizing a continuous family of finger solutions. However, experiments only showed fingers with
$\lambda=1/2$. This selection problem was understood only three decades later by different groups
\cite{shra86,hong86,comb86,tanv87,kess88}. They included surface tension into the boundary condition for the
pressure $p$ on the interface. This boundary condition also prevents cusp formation within a finite time
\cite{shra84,howi86}; this leads to a regularized moving boundary problem. It was shown, using expansion beyond all orders and reduction to a nonlinear eigenvalue
problem, that in the limit of small surface tension only the finger with $\lambda=1/2$ is stable. Recently it
was found that the so-called ``kinetic undercooling'' boundary condition also leads to regularization and
dynamical selection of the Saffman-Taylor finger with width $\lambda=1/2$ for infinitesimally weak
regularization~\cite{chap03}. We recently have proposed a similar regularization mechanism for
streamers~\cite{meul05,brau07}.

\begin{figure}
\includegraphics[width=\columnwidth]{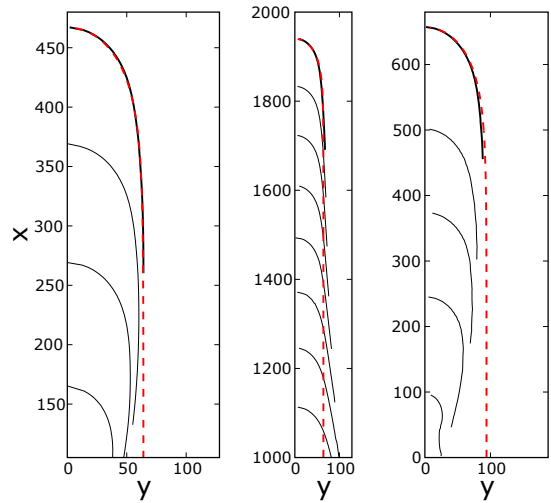}
\caption{Solid lines: contour lines characterizing the simulations at time steps
of $\Delta t = 100$; dashed curves: the uniformly translating Saffman-Taylor finger solution
(\ref{steq}) with $\lambda=1/2$.
$(a)$ background electric field $E_{\infty} = 0.5$
and width $L = 256$ with an initial seed smaller than the steady state solution (same as in Fig.~\ref{chargedens}),
$(b)$ same as in (a) but with an initial seed wider than the asymptotic solution (here it takes
longer to reach uniform translation) and $(c)$ $E_{\infty} = 0.6$, $L = 376$.
\label{fronts}}
\end{figure}

We therefore have superimposed the Saffman-Taylor finger with width $L/2$ as a solid line on
the streamer in Fig.~\ref{chargedens}. The agreement is convincing. In Fig.~\ref{fronts},
the comparison is further elaborated for three different cases, where $(b)$ differs from $(a)$
by the initial density distribution, and $(c)$ by $E_{\infty}$ and $L$.
Here solid lines represent stages of evolution of the density model from initial transients
to uniform translation; they indicate the position $y(x)$ of the maximal charge density for every $x$.
The dashed lines are
the Saffman-Taylor solution (\ref{steq}) with the selected width $\lambda=1/2$. No adjustments are possible,
except for an arbitrary translation of the Saffman-Taylor finger along the $x$-axis.  This is chosen to overlap with the latest stage of the density evolution which is the attractor of the dynamics (at a later stage the shape of the front stays identical and it moves at constant speed). Again the agreement is very convincing. A direct consequence of this agreement
is that we expect the field to be enhanced by a factor of 2 immediately ahead of the front, and the finger velocity to be $2E_{\infty}$, independently of the values of $L$ and $E_{\infty}$.
Indeed we observe that this value of the enhanced field is reached when the moving boundary approximation is most accurate, i.e. when
the width of the space-charge layer is much smaller than the radius of curvature of the front.  The former is rather independent on $L$ and $E_{\infty}$, while the latter is of order $L$, since the width of
streamer approaches $L/2$.  However, since large $L$ also leads to branching, this behavior is observed only for parameters slightly to the left of the phase-separation curve of Fig.~\ref{phasediagram}.


\subsection{Open problems for boundary analysis}
\label{sec:profile-c}

However, this apparently very successful interfacial model relies on four approximations. 
\begin{enumerate}
\item For the front
motion, the electron drift velocity $v$ in the local electric field $E$ is increased by a diffusion-reaction
correction~\cite{eber96}. The present simulations show that the streamer velocity in the maximal electric
field $E^+$ can be linearly interpolated by $v=1.312\, E^+ + 6\cdot 10^{-4}$ within the explored field range,
giving values closely below the analytical result $v=|E| + 2\sqrt{D |E| e^{-1/|E|}}$ for planar fully relaxed fronts \cite{eber96}. Such a velocity correction $v=c\;E^+$ can be absorbed completely into rescaling
time with $c$. 
\item The streamer interior is not field free immediately behind the ionization front as
Fig.~\ref{profile} shows. Consequently, in contrast to the prediction of the moving boundary approximation the front obtained from the minimal streamer model (\ref{pde1})-(\ref{pde3}) is not completely equipotential, as Fig.~\ref{chargedens} shows.
\item The space charge layer has a finite width. As a consequence, it also can
be seen in this figure that the electric field is not enhanced by a factor of 2 but somewhat less, while
the interface position agrees very well. 
\item The interfacial approximation breaks down at the sides of
the streamer finger where the local electric field {\bf E} is too low to sustain substantial ionization,
$e^{-1/|{\bf E}|}\ll 1$, while the interface between two fluids in the Saffman-Taylor finger, of course,
continues along the whole channel length.  Therefore the mathematical similarity between Saffman-Taylor
fingers and streamer fingers holds only close to their tips, while the analytical construction of fingers
requires their whole length. 
\end{enumerate}

These observations pose new challenges to the theoretical understanding of finger selection in moving boundary problems.


\section{Summary and conclusions}
\label{sec:conclusions}
This paper presents, up to our knowledge, the first studies on the full
dynamics of multiple interacting streamers.  By using a simplified but 
physically relevant model, we were able to focus on the main effects
of the interaction and stress the most general electro-dynamic properties
of a bunch of streamers. We obtained a phase diagram spanned by the electric field $E_{\infty}$ and the spatial period $L$, see Fig.~\ref{phasediagram}. For $L$ and/or $E_{\infty}$ large enough, the streamers branch similarly to single streamers. For $L$ and/or $E_{\infty}$ small enough, the streamers do not branch and approach the width $L/2$. Furthermore, we used a moving boundary approximation to derive surprisingly accurate predictions. We showed that close to the braching line of the phase diagram, the enhanced field at the tip of the streamer is close to $2 E_{\infty}$, where $E_{\infty}$ is the background electric field applied between the electrode. Moreover we showed that the shape of the front is well fitted by the selected Saffman-Taylor finger derived analytically from the moving boundary approximation.

Certainly there are still many open questions about this topic.
Further investigations should extend our model to three spatial
dimensions and to a wider variety of
media, including nonlocal ionization mechanisms~\cite{luque07}.
A rigorous analysis of the problem of finger selection in this context
of interacting streamers would also prove valuable both for the
pattern formation community and for an improved understanding of streamers.

\begin{acknowledgments}
A.L.\ was supported by STW--project 06501, and F.B.\ by project 633.000.401 within the program ``Dynamics of
Patterns'', both within the Netherlands Organization for Scientific Research NWO.
\end{acknowledgments}


\begin{thebibliography}{99}

\bibitem{raiz91} Y.P. Raizer, {\it Gas Discharge Physics}, Springer-Verlag, Berlin, 1991.

\bibitem{veld99} E.M. van Veldhuizen (ed.), {\it Electrical Discharges for Environmental Purposes:
Fundamentals and Applications}, NOVA Science Publishers, New York, 1999.

\bibitem{eber06} U. Ebert {\it et al.}, Plasma Sources Sci. Technol. {\bf 15}, S118 (2006).

\bibitem{sent95} D.D. Sentman, Geophys. Res. Lett. {\bf 22}, 1205 (1995).

\bibitem{pask98} V.P. Pasko {\it et al.}, Geophys. Res. Lett. {\bf 25}, 2123 (1998).

\bibitem{Gerken} E.A. Gerken, U.S. Inan, C.P. Barrington-Leigh, Geophys. Res. Lett. {\bf 27}, 2637 (2000).

\bibitem{pask02} V.P. Pasko {\it et al.}, Nature (London) {\bf 416}, 152 (2002).

\bibitem{niel07} M.G. McHarg {\it et al.}, Geophys. Res. Lett. {\bf 34}, L06804 (2007).

\bibitem{Winands06} G. J. J. Winands {\it et. al}, J. Phys. D: Appl. Phys. {\bf 39}, 3010 (2006).

\bibitem{Briels06} T. M. P. Briels {\it et. al} J. Phys. D: Appl. Phys. {\bf 39}, 5201 (2006).

\bibitem{naid96} G.V. Naidis, J. Phys. D: Appl. Phys. {\bf 29}, 779 (1996).

\bibitem{takak05} K. Takaki {\it et al.}, Appl. Phys. Lett. {\bf 86}, 151501 (2005).

\bibitem{stari05}  A. V Krasnochub {\it et. al}, Proceedings of the XXVIIth
ICPIC, Eindhoven, The Netherlands, No. 04-312 (2005).

\bibitem{pelc88} P. Pelce, {\it Dynamics of Curved Fronts}, Academic Press, New York, 1988.

\bibitem{saff58} P.G. Saffman, G.I. Taylor, Proc. R. Soc. London A {\bf 245}, 312 (1958).

\bibitem{Kunhardt} C. Wu and E.E. Kunhardt, Phys. Rev. A {\bf 37}, 4396 (1988).

\bibitem{dhal85} S.K. Dhali and P.F. Williams, Phys. Rev. A {\bf 31}, 1219 (1985);
J. Appl. Phys. {\bf 62}, 4696 (1987).

\bibitem{vite94} P.A. Vitello, B.M. Penetrante, and J.N. Bardsley, Phys. Rev. E {\bf 49}, 5574 (1994).

\bibitem{eber96} U. Ebert, W. van Saarloos, and C. Caroli, Phys. Rev. Lett. {\bf 77}, 4178 (1996);
Phys. Rev. E {\bf 55}, 1530 (1997).

\bibitem{baze98} E.M. Bazelyan and Yu.P. Raizer, Spark Discharges (CRS Press, New York, 1998).

\bibitem{luque07} A. Luque {\it et al.}, Appl. Phys. Lett. {\bf 90}, 081501 (2007).

\bibitem{mon2006} C. Montijn {\it et al.}, J. Comput. Phys. {\bf 219}, 801 (2006).

\bibitem{brau07} F. Brau {\it et al.}, Phys. Rev. E {\bf 77}, 026219 (2008).

\bibitem{Schoell} G. Schwarz {\it et al.},
Physica B {\bf 272}, 270 (1999); Phys. Rev. B {\bf 61}, 10194 (2000);
Semicond. Sci. Technol. {\bf 15}, 593 (2000);  Physica E {\bf 12}, 182 (2002).

\bibitem{arra02} M. Array\'as {\it et al.},
Phys. Rev. Lett. {\bf 88}, 174502 (2002).

\bibitem{monbr} C. Montijn {\it et al.},
Phys. Rev. E {\bf 73}, 065401 (2006).

\bibitem{meul04} B. Meulenbroek {\it et al.}, Phys. Rev. E {\bf 69}, 067402 (2004).

\bibitem{meul05} B. Meulenbroek {\it et al.}, Phys. Rev. Lett. {\bf 95}, 195004 (2005).

\bibitem{eber07} U. Ebert, B. Meulenbroek, L. Sch\"afer, SIAM J. Appl. Math. {\bf 69}, 292 (2007).

\bibitem{shra86} B.I. Shraiman, Phys. Rev. Lett. {\bf 56}, 2028 (1986).

\bibitem{hong86} D.C. Hong, J.S. Langer, Phys. Rev. Lett. {\bf 56}, 2032 (1986).

\bibitem{comb86} R. Combescot {\it et al.}, Phys. Rev. Lett. {\bf 56}, 2036 (1986).

\bibitem{tanv87} S. Tanveer, Phys. Fluid {\bf 30}, 1589 (1987).

\bibitem{kess88} D. Kessler {\it et al.}, Adv. Phys. {\bf 37}, 255 (1988).

\bibitem{shra84} B. Shraiman, D. Bensimon, Phys. Rev. A {\bf 30}, 2840 (1984).

\bibitem{howi86} S.D. Howison, SIAM J. Appl. Math. {\bf 46}, 20 (1986).

\bibitem{chap03} S.J. Chapman, J.R. King, J. Eng. Math. {\bf 46}, 1 (2003).


\end{thebibliography}
\end{document}